      [1999/12/01 v1.4c Il Nuovo Cimento]
\documentclass{cimento}
\title{Strangelets at Chacaltaya} 
\author{M.~Rybczy\'nski\from{ins:K}\thanks{e-mail: mryb@pu.kielce.pl}\ETC,
Z.~W\l odarczyk\from{ins:K}\thanks{e-mail: wlod@pu.kielce.pl},
        \atque
G.~Wilk\from{ins:W}\thanks{e-mail: wilk@fuw.edu.pl}}
\instlist{\inst{ins:K} Institute of Physics, Pedagogical University,
                       Kielce, Poland 
  \inst{ins:W} The Andrzej Soltan Institute for Nuclear Studies,
               Warsaw, Poland  }
\PACSes{\PACSit{96.40}{Cosmic rays}
\PACSit{01.30.Cc}{Conference proceedings}}
\begin{document}

\maketitle

\begin{abstract}
We discuss the possible imprints of strangelets (i.e., lumps of
Strange Quark Matter) in Chacaltaya experimental data using model of
propagation of such objects through the atmosphere developed by us
recently.
\end{abstract}

\section{Introduction: strangelets and their propagation}

Chacaltaya Laboratory offers unique possibility to observe possible
imprints of stran\-ge\-lets arriving from the outer space. They are
lumps of Strange Quark Matter (SQM), a new possible stable form of
matter (cf.
\cite{ref:astro,ref:sqm,ref:greek,ref:af,ref:stab,ref:kas} for
details). Following \cite{ref:ww} it is fully sensible to search for
strangelets in cosmic ray experiments, especially at Chacaltaya
level ($540$ g/cm$^{2}$ of atmosphere) because \cite{ref:ww} the
specific features of strangelets \cite{ref:af} allow them to
penetrate deep into atmosphere \cite{ref:wwk}. The point is that there
exists some critical size of strangelet given by the critical value of
its mass number $A = A_{crit} \sim 300\div 400$ such that for 
$A > A_{crit}$ strangelets are absolutely stable against neutron
emission. (However, small strangelets might probably also gain
stability due to the shell effect \cite{ref:stab}.). Below this limit
strangelets decay rapidly evaporating neutrons. The geometrical radii
of strangelets turn out to be comparable to the radii of ordinary
nuclei \cite{ref:ww}, i.e., their geometrical cross sections are
similar to the normal nuclear ones. To account for their strong
penetrability one has to accept  that strangelets reaching deeply into
atmosphere are formed in many successive interactions with air nuclei
by the initialy very heavy lumps of SQM entering the atmosphere and
decreasing due to the collisions with air nuclei (until their $A$ 
reaches the critical value $A_{crit}$ \cite{ref:ww}). The opposite
scenario advocated recently in \cite{ref:sr} faces some difficulties
and will not be discussed here.

Such scenario is fully consistent with all present experiments
\cite{ref:ww}. In this scenario  interaction of strangelet with 
target nucleus involves all quarks of the target located in 
the geometrical intersection of the colliding air nucleus and
strangelet. It is assumed that each quark from the target interacts
with only one  quark from the strangelet; i.e., during interaction
the mass  number of strangelet is diminished to the value equal to
$A-A_t$ at  most. This procedure continues unles either strangelet
reaches  Earth or (most probably) disintegrates at some depth $h$ of the 
atmosphere reaching $A(h)=A_{crit}$. This results, in a first 
approximation (in which $A_t << A_{crit} < A_0$), in total
penetration depth of the order of  $\Lambda \simeq \frac{4}{3} 
\lambda_{NA_t}(A_0/A_t)^{1/3}$. The characteristic features of
strangelets propagation are illustrated in Fig. 1. All numerical
calculations presented here were done using suitable modifications
of the SHOWERSIM \cite{ref:num} modular  software. Strangelets  
propagation and nuclear-electromagnetic cascades through the 
atmosphere were simulated with primaries ($p$, $Fe$ and 
strangelets) initiating showers sampled from the 
power spectrum $F(E) \sim E^{-2.7}$ with energies above $1000$ TeV
per particle. EAS detected at Chacaltaya with $N_e = 10^6 \div 10^7$
were then analysed. 

\begin{figure}[h]
\setlength{\unitlength}{1cm}
\begin{picture}(14,8)
\includegraphics{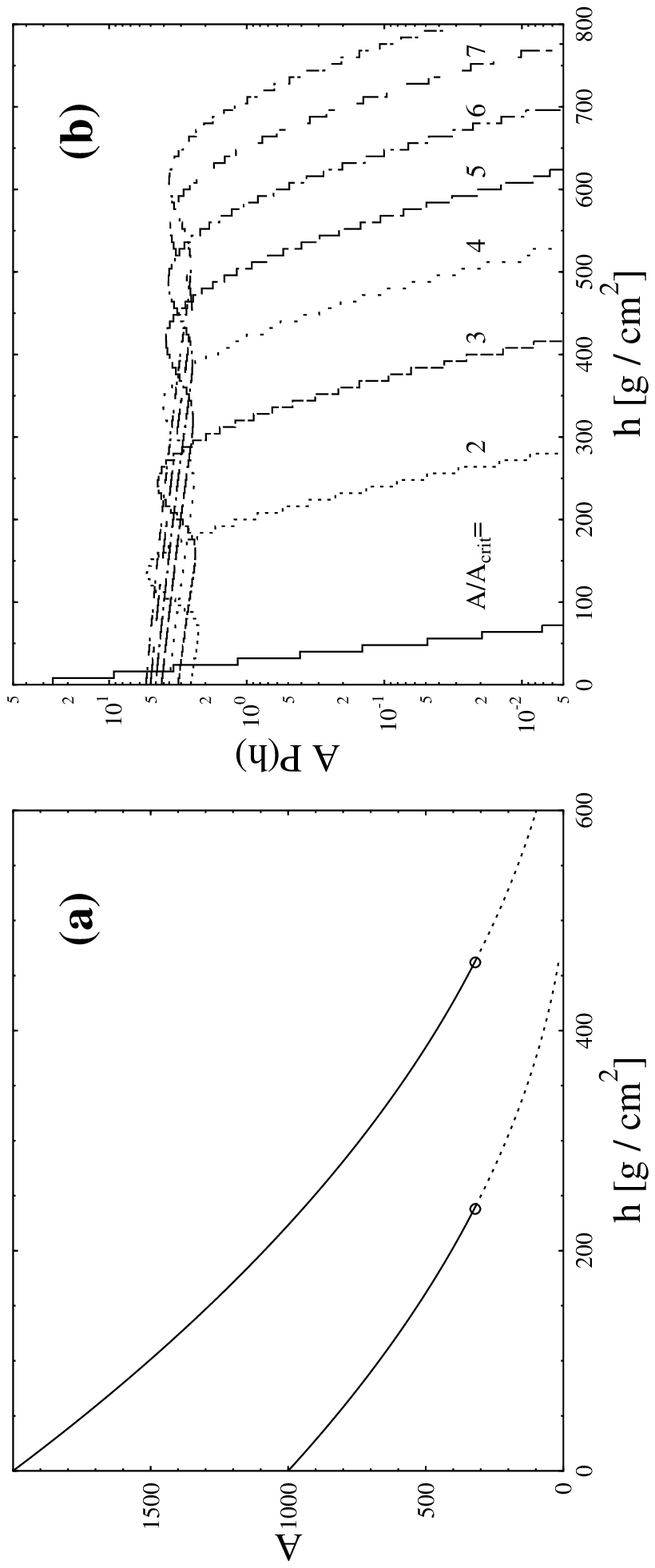}
\end{picture}
\vspace{-0.5cm}
\end{figure}
\vspace{-2.1cm}
\begin{minipage}[h]{12.9cm}
\noindent
Fig. 1. $(a)$ An example of the predicted decrease of the actual size 
of strangelet $A$ with depth $h$ of the atmosphere traversed for two 
different initial sizes: $A_0 = 1000$ and $2000$ 
(dotted lines correspond to $A < A_{crit}$).
(b) Number of nucleons released in $1$ g/cm$^2$ at depth $h$ of the 
atmosphere from the strangelet with mass number ratios $A_0/A_{crit} = 1,~ 
2,\dots,~ 8$, respectively.
\end{minipage}
\vspace{2mm}

\section{Cosmic nuclearities and exotic events}

There are several reports suggesting existence of direct candidates
for SQM \cite{ref:examples} (characterized mainly by their very small
ratios of $Z/A$). All of them have mass numbers $A$ near or slightly
exceeding $A_{crit}$ (including Centauro event which contains
probably  $\sim 200$ baryons \cite{ref:b}). Analysis of these 
candidates for SQM shows \cite{ref:ww} that the abundance of strangelets in the
primary cosmic ray flux is $F_S(A_0=A_{crit})/F_{tot} \simeq 2.4
\cdot 10^{-5}$ at the same enery per particle. Efficiency for 
registration of strangelets at Chacaltaya level is shown in Fig. 2a.
To detect strangelets with $A>A_{crit}$ at Chacaltaya the mass of the
initial strangelet should be $A_0 \simeq 7 A_{crit}$ what leads to
$\sim 10^{-11}$ as the relative  abundance of such strangelets.
For normal flux of primary cosmic rays \cite{ref:shib} the
expected flux of strangelets is then equal to $F_S = 7\cdot 10^{-6}$
m$^{-2}$h$^{-1}$sr$^{-1}$
for the energy  above $10$ GeV per initial strangelet. The high altitude
exposure of  passive nuclear track detector arrays \cite{ref:bak} 
and their operation for  one year allow therefore detection of such 
objects. In fact, the exposed CR39 detector should detect at least $6$
strangelets with masses  above $A_{crit}$ and with energies per registered
strangelet above $1$ GeV. The  mass distribution of strangelets,
which can be observed at Chacaltaya  is shown in Fig. 2b.

\begin{figure}[h]
\setlength{\unitlength}{1cm}
\begin{picture}(14,7.8)
\includegraphics{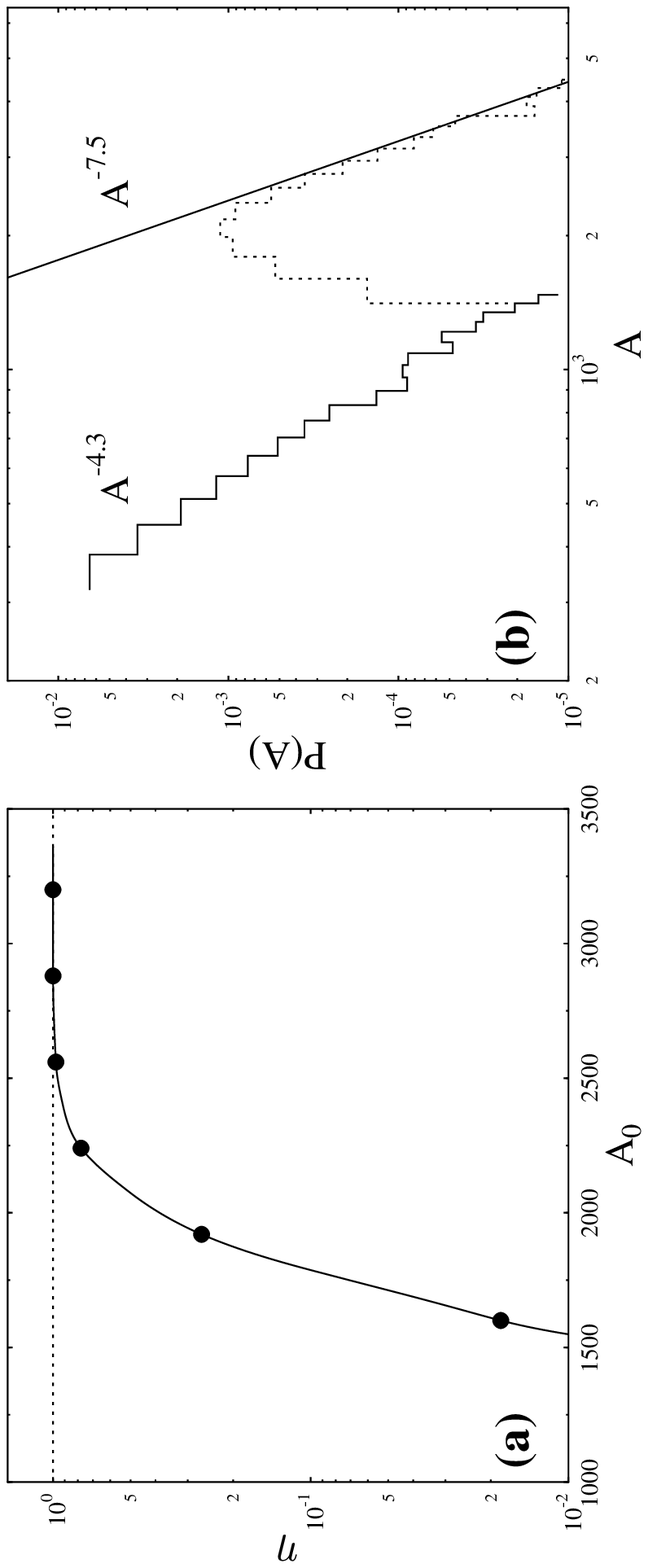}
\end{picture}
\vspace{-0.5cm}
\end{figure}
\vspace{-2.75cm}
\begin{minipage}[h]{12.9cm}
\noindent
Fig. 2. $(a)$ Registration eficiency for strangelet with mass 
number $A > A_{crit}$ at the Chacaltaya level as funcion of initial 
mass number $A_0$. Consecutive full circles indicate (for $A_0 > 1600$) 
points where $A_0/A_{crit} =  5,~ 6,~ 7,~ 8,~ 9,~ 10$, respectively.
$(b)$ Mass distribution of strangelets at Chacaltaya level (solid 
histogram) resulting from primary mass spectrum $A^{-7.5}$ (solid line). 
The corresponding initial mass distribution for detected strangelets 
is shown by dotted histogram.
\end{minipage}
\vspace{2mm}

Experimental results obtained at Chacaltaya show a wide spectrum of
exotic events (Centauros, superfamilies with 'halo', strongly
penetrating component, etc.) which are clearly incompatible with
the standard ideas of hadronic interactions known from the accelerator
experiments. Some new mechanism or new primaries are therefore
needed. Assuming that strangelets represent such new primaries one is
able to explain \cite{ref:gdw} (at least to some extend) a strong
penetrating nature of some `abnormal' cascades associated with their
very slow attenuation and with the apperance of many maxima with small
distances between them (about $2-3$ times smaller than in the
'normal' hadron cascades).

Already mentioned Centauro (and mini-Centauro) events, characterized
by the extreme imbalance between hadronic and gamma-ray components
among produced secondaries, are probably the best known examples
of such exotic events. They require deeply penetrating component in
cosmic rays. We claim that they can be products of strangelets
penetrating deeply into atmosphere and evaporating neutrons
\cite{ref:wwk}. Both the flux ratio of Centauros registered  at different
depths and the energy distribution within them can be  successfully
described by such concept. 

Another example of exotic event is phenomenon of alignment of
structural objects of gamma-hadron families near a stright line in
the plane at the target diagram \cite{ref:boris}. The excess of
aligned families observed is incompatible with any conventional
concept of interaction. One can speculate therefore that it is 
caused by the arrival of strangelets with high spin ($J\sim A^2$ )
gradually  dispersing their masses $A(h)$ when propagating through
the atmosphere. 

Anomalous events have been reconfirmed by measuring extensive air
showers (EAS) \cite{ref:eas}. Among them was the striking observation
\cite{ref:neutron} of extremely long-delayed neutrons in conjection
with the large EAS which can not be explained by the known mechanism of
hadronic cascades development. Also muon bundles of extremaly 
high multiplicity observed recently 
by ALEPH detector (in the dedicated cosmic-ray run) can orginate 
from strangelets collisions with the atmosphere \cite{ref:aleph}.
As an illustration of sensitivity of EAS characteristics on primary 
strangelets we shown in Fig. 3 our predictions the corresponding 
distributions of hadrons and muons in EAS detected at Chacaltaya.

\begin{figure}[h]
\setlength{\unitlength}{1cm}
\begin{picture}(14,8)
\includegraphics{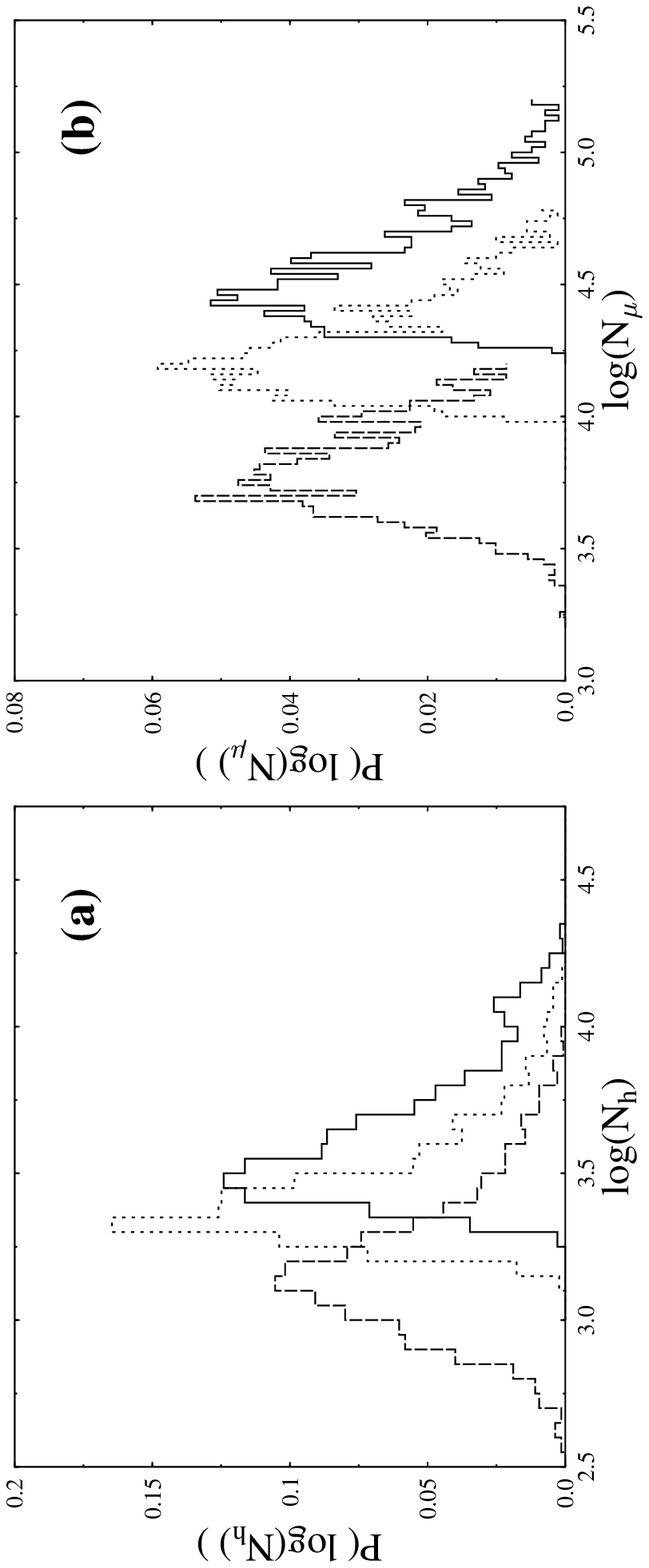}
\end{picture}
\vspace{-0.5cm}
\end{figure}
\vspace{-2.7cm}
\begin{minipage}[h]{12.9cm}
\noindent
Fig. 3. Multiplicity distribution of $(a)$ hadrons and 
$(b)$ muons in EAS with size $N_e=10^6 \div 10^7$ 
detected at Chacaltaya and initiated by primary protons 
(dashed), iron nuclei (dotted) and strangelets with 
$A_0=400$ (solid histogram).
\end{minipage}
\vspace{2mm}

\begin{figure}[h]
\setlength{\unitlength}{1cm}
\begin{picture}(14,8)
\includegraphics{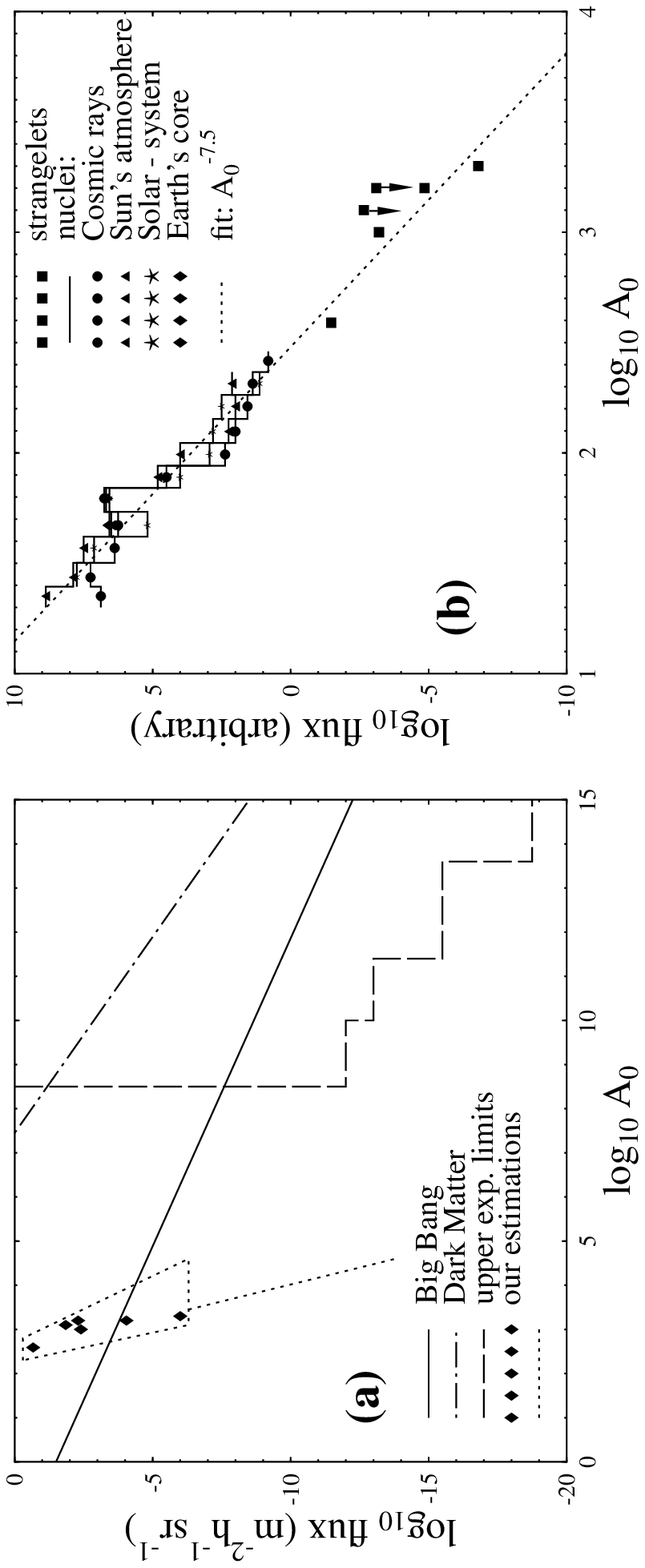}
\end{picture}
\vspace{-0.5cm}
\end{figure}
\vspace{-2.9cm}
\begin{minipage}[h]{12.9cm}
\noindent
Fig. 4. $(a)$ The expected flux (our results) of strangelets 
compared with the upper experimental limit, compiled by Price \cite{ref:limit}, 
and predicted astrophysical limits: Big Bang estimation comes 
from nucleosynthesis with quark nuggets formation; Dark Matter 
one comes from local flux assuming that galactic halo density 
is given solely by quark nuggets.
$(b)$ Comparision of the estimated mass spectrum $N(A_0)$ for 
strangelets with the known abundance of elements in the Universe 
\cite{ref:zdanow}. 
\end{minipage}
\vspace{2mm}

\section{Final remarks}

The experimental data mentiond before lead to the flux of strangelets
which is consistent (cf. Fig. 4a) with the astrophysical limits and
with the upper limits given experimentally \cite{ref:limit}. It
follows the $A_0^{-7.5}$ behaviour, which coincides with the
behaviour of abundance of normal nuclei in the Universe (Fig. 4b)
\cite{ref:zdanow}. The fascinating subject of searches for
strangelets as a new form  of matter can be succesfully realised at
experiments located in  the mountain region of Chacaltaya.
Interpretation of indirect  observations (anomalous events observed
in emulsion chambers  and results from the measurements of EAS) can provide
signals of strangelets.  Moreover, direct identification (by implementing
passive nuclear  track detector arrays) of SQM is quite realistic in
the near future.  All these justifies interest in further
experimental  search for the SQM and for its cosmological and 
elementary particle physics aspects.

\acknowledgments
Authors ZW and GW are gratefull to the organizers of the {\it
Chacaltaya Meeting on Cosmic Ray Physics} for their support and
hospitality. The partial support of Polish Committee for 
Scientific Research (grants 2P03B 011 18 and 
621/E-78/SPUB/CERN/P-03/DZ4/99) is acknowledged.

\end{document}